\documentclass[a4paper]{llncs}

\usepackage{latexsym}
\usepackage{amssymb}
\usepackage{amsmath}
\usepackage{frameit}
\usepackage{graphicx,subfigure}
\usepackage{pstricks,pst-node,pst-tree}
\usepackage{mathpartir}
\usepackage{haskell}
\usepackage{relsize}

\input xy

\xyoption{all} 

\usepackage{listings}
\lstnewenvironment{code}
    {\lstset{}%
      \csname lst@SetFirstLabel\endcsname}
    {\csname lst@SaveFirstLabel\endcsname}
\newcommand{\bi}{\begin{array}[t]{@{}l@{}}}
\newcommand{\ei}{\end{array}}
\newcommand{\ba}{\begin{array}}
\newcommand{\ea}{\end{array}}
\newcommand{\bda}{\[\ba}
\newcommand{\eda}{\ea\]}
\newcommand{\bp}{\begin{quote}\tt\begin{tabbing}}
\newcommand{\ep}{\end{tabbing}\end{quote}}

\newcommand{\tlabel}[1]{\mbox{(#1)}}

\newcommand{\myirule}[2]{{\renewcommand{\arraystretch}{1.2}\ba{c} #1
                      \\ \hline #2 \ea}}

\newcommand{\turns}{\, \vdash \,}

\newcommand{\sgap}{\quad}

\newcommand{\mathem}{\sf}

\newcommand{\CASE}{\mbox{\mathem case}}

\newcommand{\OF}{\mbox{\mathem of}}

\newcommand{\fs}[1]{\mathit {#1}}

\newcommand{\canonic}[1]{{\cal C}(#1)}
\newcommand{\injs}[2]{{\mathit injs}_{#1} \ #2}
\newcommand{\allParse}[2]{{\mathit allParse} \ #1 \ #2}
\newcommand{\simp}[3]{#1 \stackrel{\mathit{#2}}{\gg} #3}

\newcommand{\arrow}{\rightarrow}

\newcommand{\comment}[1]{}
\newcommand{\ignore}[1]{}
\newcommand{\ms}[1]{{\bf MS:#1}}       
\newcommand{\pjs}[1]{}
\newcommand{\jw}[1]{}

\newcommand{\mysection}[1]{\vspace*{-2mm}\section{#1}\vspace*{-1mm}}

\newcommand{\lang}{{\cal L}}

\newcommand{\deriv}[2]{d_{#2}(#1)} 
\newcommand{\pderiv}[2]{pd_{#2}(#1)}

\newcommand{\allEps}[1]{{\mathit allEps}_{#1}}

\newcommand{\conc}{\cdot}

\newcommand\Descendant\preceq

\newcommand{\flatten}[1]{|#1|}

\newcommand{\Left}{\mathit{L}}
\newcommand{\Right}{\mathit{R}}

\newcommand{\AllParseFST}[1]{{\cal FST}(#1)}

\newcommand{\Desc}[1]{{\cal D}(#1)}

\title{Derivative-Based Diagnosis of Regular Expression Ambiguity}

\author{Martin Sulzmann\inst{1} and Kenny Zhuo Ming Lu \inst{2}}
\institute{Karlsruhe University of Applied Sciences \\
  \email{martin.sulzmann@hs-karlsruhe.de}
  \and
   Nanyang Polytechnic \\
  \email{luzhuomi@gmail.com}}

\begin{document}

\maketitle

\begin{abstract}
  Regular expressions are often ambiguous.
  We present a novel method based on Brzozowski's derivatives
  to aid the user in diagnosing ambiguous regular expressions.
  We introduce a derivative-based finite state transducer to generate parse trees
  and minimal counter-examples.
  The transducer can be easily customized
  to either follow the POSIX or Greedy disambiguation policy
  and based on a finite set of examples it is possible to
  examine if there are any differences among POSIX and Greedy.
  \\ \mbox{} \\
  \noindent{\bf Keywords:} regular expressions, derivatives, ambiguity, POSIX, Greedy
\end{abstract}

\mysection{Introduction} \label{sec:intro}


A regular expression is \emph{ambiguous} if a string
can be matched in more than one way.
For example, consider the expression $x^* + x$ where
input string $x$ can either be matched against
$x^*$ or $x$. Hence, this expression is ambiguous.



\paragraph{Earlier works}

There exist well-established algorithms to check
for regular expression ambiguity. However,
most works report ambiguity in terms of an automata which
results from an ambiguity-preserving translation
of the original expression,
e.g. see the work by Book, Even, Greibach and Ott~\cite{Book:1971:AGE:1309278.1309583}.
From a user perspective, it is much more useful to report
ambiguity in terms of the original expression.
We are only aware of two works which like us perform the
ambiguity analysis on the original expression.

Brabrand and Thomsen~\cite{DBLP:conf/ppdp/BrabrandT10}
establish a structural relation to detect ambiguity
based on which they can provide minimal counter examples.
They consider some disambiguation strategies but do not
cover the POSIX interpretation.

Borsotti, Breveglieri, Crespi{-}Reghizzi and Morzenti \cite{DBLP:conf/wia/BorsottiBCM15}
show how to derive parse trees based on marked regular expressions~\cite{mnya60} as employed
in the Berry-Sethi algorithm~\cite{Berry:1986:RED:39528.39537}.
They establish criteria to identify ambiguous regular expressions.
Like ours, their approach can be customized to support either
the POSIX~\cite{posix} or Greedy~\cite{pcre} disambiguation policy.
However, for POSIX/Greedy disambiguation,
their approach requires tracking of dynamic data based
on the Okui-Suzuki method~\cite{Okui:2010:DRE:1964285:1964310}.
Our approach solely relies on derivatives, no dynamic tracking of data is
necessary.

\paragraph{Our work}

Brzozowski's derivatives~\cite{321249} support the symbolic construction of automata
where expressions represent automata states.
This leads to elegant algorithms and based on
(often simple) symbolic reasoning.
In earlier work~\cite{DBLP:conf/flops/SulzmannL14},
we have studied POSIX matching based on derivatives.
In this work, we show how to adapt and extend
the methods developed in~\cite{DBLP:conf/flops/SulzmannL14}
to diagnose ambiguous expressions. 



\paragraph{Contributions and outline}

In summary,  our contributions are:
\begin{itemize}
  \item We employ derivatives to compute all parse trees
        for a large class of (non-problematic) regular
        expressions (Section~\ref{sec:all-match}).
  \item We can build a finite state transducer to compute
    these parse trees (Section~\ref{sec:all-match-fsa}).
  \item We can easily customize the transducer
    to either compute the POSIX or greedy parse tree
      (Section~\ref{sec:posix-and-greedy}).
  \item We can identify
        simple criteria 
        to detect ambiguous expressions
        and to derive a finite set of minimal counter-examples.
        Thus, we can statically verify if there are any differences
        among POSIX and Greedy 
        (Section~\ref{sec:diagnos-ambig}).
      \item We have implemented the approach in Haskell.
        The implementation is available via
         \verb+http://www.home.hs-karlsruhe.de/~suma0002/dad.html+.
\end{itemize}

In the upcoming section, we introduce our notion of regular expression,
parse trees and ambiguity. 

\emph{Proof sketches for results stated can be found in the appendix.}

\mysection{Regular Expressions, Parse Trees and Ambiguity}  
\label{:sec:regex}

The development largely
follows~\cite{cduce-icalp04}
and \cite{DBLP:conf/ppdp/BrabrandT10}.
We assume that symbols are taken from a fixed, finite
alphabet $\Sigma$. We generally write $x, y, z$
for symbols. 

\begin{definition}[Words and Regular Expressions]
Words are either empty or
concatenation of words and defined as follows:  
$ w  ::=  \epsilon \mid x \in \Sigma \mid w \conc w $.
  
We denote regular expressions by $r,s,t$.
Their definition is as follows:
$
 r  ::= x \in \Sigma \mid r^* \mid r \conc r \mid r + r \mid \epsilon \mid \phi
$
The mapping to words is standard.
$\lang(x) = \{ x \}$.
$\lang(r^*) = \{ w_1 \conc ... \conc w_n \mid n \geq 0 \wedge 
                                w_i \in \lang(r) \wedge
                                i \in \{1,..,n\} \}$.
$\lang(r \conc s) = \{ w_1 \conc w_2 \mid w_1 \in \lang(r) \wedge w_2 \in \lang(s) \}$.
$\lang(r + s) = \lang(r) \cup \lang(s)$.
$\lang(\epsilon) = \{ \epsilon \}$.
$\lang(\phi) = \{ \}$.

We say an expression $r$ is \emph{nullable}
iff $\epsilon \in \lang(r)$.
\end{definition}
As it is common, we assume that $+$
and $\conc$ are right-associative.
That is, $x + y + x\conc y \conc z$
stands for $x + (y + (x \conc (y \conc z)))$.

A parse tree explains
which subexpressions match which subwords.
We follow~\cite{cduce-icalp04} and view
expressions as types and parse trees as values.

\begin{definition}[Parse Trees]
  Parse tree values are built using
  data constructors such as lists, pairs, left/right injection into a disjoint sum etc.
In case of repetitive matches such as in case of Kleene star,
we make use of lists.
We use Haskell style notation
and write $[v_1,...,v_n]$ as a short-hand
for $v_1 : ... : v_n : []$.
\bda{c}
 v \ ::= \ () \mid x \mid (v,v)
 \mid  \Left~v \mid \Right~v
    \mid  vs 
 \ \ \ \ 
vs \ ::= \ [] \mid v : vs
\eda
The valid relations among parse trees and regular expressions
are defined via a natural deduction style
proof system.
\bda{c}
\turns [] : r^*
 \ \ \ \ 
\myirule{\turns v : r
         \sgap \turns vs : r^*}
        {\turns (v:vs) : r^*}
 \ \ \ \
\myirule{ \turns v_1 : r_1 \sgap \turns v_2 : r_2 }
         {\turns (v_1, v_2) : r_1 \conc r_2 }
\\ 
 \myirule{\turns v_1 : r_1}
         {\turns \Left~v_1 : r_1 + r_2}
 \ \ \ \
 \myirule{\turns v_2 : r_2}
         {\turns \Right~v_2 : r_1 + r_2}
 \ \ \ \        
 \turns () : \epsilon
 \ \ \ \
 \myirule{x \in \Sigma}
         { \turns x : x }
\eda
\end{definition}

\begin{definition}[Flattening]
  We can flatten a parse tree to a word as follows:
\bda{lllllllllllllllll}
 \flatten{()}  &=&  \epsilon &
  \flatten{x} &=& x          &
  \flatten{\Left~v}  &=&  \flatten{v} &
  \flatten{v:vs}  &=&  \flatten{v} \conc \flatten{vs}
\\
 \flatten{[]} & =&  \epsilon  ~~~&
 \flatten{(v_1,v_2)} &=& \flatten{v_1} \conc \flatten{v_2} ~~~&
 \flatten{\Right~v} &=&  \flatten{v}   ~~~ &
\eda
\end{definition}

\begin{proposition}[Frisch/Cardelli~\cite{cduce-icalp04}]
Let $r$ be a regular expression.
If $w \in \lang(r)$ for some word $w$, then there
exists a parse tree $v$ such
that $\turns v : r$ and $\flatten{v} = w$.
If $\turns v : r$ for some parse tree $v$,
then $\flatten{v} \in \lang(r)$.
\end{proposition}

\begin{example}
  We find that $x \conc y\in \lang((x\conc y + x + y)^*)$
where $[\Left\ (x,y)]$ is a possible parse tree.
Recall that $+$ is right-associative and therefore
we interpret $(x\conc y + x + y)^*$
as $(x\conc y + (x + y))^*$.
\end{example}

An  expression is ambiguous if there exists
a word which can be matched in more than one way.
That is, there must be two distinct parse trees
which share the same underlying word.

\begin{definition}[Ambiguous Regular Expressions]
We say a regular expression $r$ is \emph{ambiguous} iff
there exist two distinct parse trees $v_1$ and $v_2$ such
that $\turns v_1 : r$ and $\turns v_2 : r$ 
where $\flatten{v_1} = \flatten{v_2}$.
\end{definition}

\begin{example}
  $[\Left \ (x,y)]$ and $[\Right \ (\Left\ x), \Right \ (\Right\ y)]$
are two distinct parse trees for expression $(x\conc y + x + y)^*$
and word $x \conc y$.
\end{example}

Our ambiguity diagnosis methods will operate on arbitrary expressions.
However, formal results are restricted
to a certain class of  'non-problematic' expressions.

\begin{definition}[Problematic Expressions]
We say an expression $r$ is \emph{problematic}
iff it contains some sub-expression of the form $s^*$ where $\epsilon \in \lang(s)$.
\end{definition}

For problematic expressions, the set of parse trees is infinite,
otherwise finite.
\begin{example}
\label{ex:problematic-expression}
Consider the problematic expression $\epsilon^*$
where for the empty input word we find the following (infinite)
sequence of parse trees $[]$, $[()]$, $[(), ()]$, ...
\end{example}

\begin{proposition}[Frisch/Cardelli~\cite{cduce-icalp04}]
For non-problematic expressions, the set of distinct parse 
trees which share
the same underlying word is always finite.
\end{proposition}


Next, we consider computational methods based on
Brzozowski's derivatives to compute parse trees.

\mysection{Computing Parse Trees via Derivatives} 
\label{sec:all-match}

Derivatives denote left quotients
and they can be  computed via a simple syntactic transformation.

\begin{definition}[Regular Expression Derivatives]
  The \emph{derivative} of expression $r$ w.r.t.~symbol $x$,
  written $\deriv{r}{x}$, is computed by induction on $r$:
\bda{c}
\deriv{\phi}{x}  =  \phi
\ \ \ \ \deriv{\epsilon}{x}  =  \phi
\ \ \ \ 
\deriv{r_1 + r_2}{x}  =  \deriv{r_1}{x} + \deriv{r_2}{x}
\ \ \ \ 
\deriv{r^*}{x}  =  \deriv{r}{x} \conc r^*
\\
\ba{ll}
   \deriv{y}{x}  =  \left \{   
   \ba{ll}
        \epsilon & \mbox{if~$x = y$}
    \\ \phi & \mbox{otherwise}
    \ea  \right.
    &
   \deriv{r_1 \conc r_2}{x}  = 
     \left \{
        \ba{ll}
           \deriv{r_1}{x} \conc r_2 + \deriv{r_2}{x} & 
               \mbox{if $\epsilon\in \lang(r_1)$}
         \\ \deriv{r_1}{x} \conc r_2 & \mbox{otherwise}
        \ea \right.    
\ea
        
\eda

The extension to words is as follows:
$\deriv{r}{\epsilon} = r$.
$\deriv{r}{x \conc w} = \deriv{\deriv{r}{x}}{w}$.

A \emph{descendant} of $r$ is either $r$ itself
or the derivative of a descendant.
We write $r \Descendant s$ to denote that $s$ is a descendant of $r$.
We write $\deriv{r}{}$ to denote the set of descendants of $r$.
\end{definition}

\begin{proposition}[Brzozowski~\cite{321249}]
For any expression $r$ and symbol $x$ we find
that $\lang(\deriv{r}{x}) = \{ w \mid x \conc w \in \lang(r)\}$.
\end{proposition}

Thus, we obtain a simple word matching algorithm by repeatedly
building the derivative and then checking if the final derivative
is nullable.
That is, $w \in \lang(r)$ iff $\epsilon \in \lang(\deriv{r}{w})$.
Nullability can easily be decided by induction on $r$. We omit
the straightforward details.

\begin{example}
 \label{ex:word-match}
  Consider expression $(x+y)^*$ and input $x \conc y$.
  We find $\deriv{(x+y)^*}{x} = (\epsilon + \phi) \conc (x+y)^*$
  and $\deriv{\deriv{(x+y)^*}{x}}{y} = (\phi + \phi) \conc (x+y)^* + (\phi + \epsilon) \conc (x+y)^*$.
  The final expression is nullable. Hence, we can conclude that $x \conc y \in \lang((x+y)^*)$.
\end{example}

Based on the derivative method, it is surprisingly easy to compute parse trees for some input word $w$.
The key insights are as follows:
\begin{enumerate}
  \item Build all parse trees for the final (nullable) expression.
  \item Transform a parse tree for $\deriv{r}{x}$ into
        a parse tree for $r$ by injecting symbol $x$ into
        $\deriv{r}{x}$'s parse tree.
        Injecting can be viewed as reversing the effect of the derivative operation.
\end{enumerate}


\begin{definition}[Empty Parse Trees]
Let $r$ be an expression. Then, $\allEps{r}$
yields a set of parse trees.
The definition of $\allEps{r}$ is by induction on $r$.
\begin{haskell}
\allEps{\epsilon} = \{ () \}
\ \ \ \
\allEps{\phi} = \{ \}
\ \ \ \
\allEps{x} = \{ \}
\\
\allEps{r^*} = \{ [] \} 
 \ \ \ \
\allEps{ r_1 \conc r_2 } = \{( v_1, v_2) \mid v_1 \in \allEps{r_1}
\wedge v_2\in \allEps{r_2} \}  
\\
\allEps{ r_1+r_2 } =  \{ \Left\ v_1 \mid v_1 \in \allEps{r_1} \}
                      \cup \{ \Right\ v_2 \mid v_2 \in \allEps{r_2} \}             
\end{haskell}
\end{definition}

If the expression is not nullable it is easy to see that
we obtain an empty set.
For nullable expressions, $\allEps{r}$ yields
empty parse trees.

\begin{proposition}[Empty Parse Trees]
 \label{prop:empty-parse-trees}
Let $r$ be a nullable expression.
Then, for any $v \in \allEps{r}$ we have that
$\turns v : r$ and $\flatten{v} = \epsilon$.
\end{proposition}

\begin{example}
  For the final (nullable) expression from Example~\ref{ex:word-match}
  we find that $\allEps{(\phi + \phi) \conc (x+y)^* + (\phi + \epsilon) \conc (x+y)^*} = \Right \ (\Right\ (), [])$.
\end{example}

For nullable, non-problematic expressions $r$, we can state that
$\allEps{r}$ yields all parse trees $v$ for $r$ where $\flatten{v} = \epsilon$.

\begin{proposition}[All Empty Non-Problematic Parse Trees]
\label{prop:allempty}
Let $r$ be a non-problematic expression such that $\epsilon \in \lang(r)$.
Let $v$ be a parse tree such that $\turns v : r$ where $\flatten{v}=\epsilon$.
Then, we find that $v \in \allEps{r}$.
\end{proposition}

The non-problematic assumption is necessary.
Recall Example~\ref{ex:problematic-expression}.

What remains is to describe how to derive parse trees for the original expression.
We achieve this by  injecting symbol $x$ into $\deriv{r}{x}$'s parse tree.

\begin{definition}[Injecting Symbols into Parse Trees]
\label{def:injection}
Let $r$ be an expression and $x$ be a symbol.
Then, $\injs{\deriv{r}{x}}$ is a function which maps
a $\deriv{r}{x}$'s parse tree to a set of parse trees of $r$.~footnote{Additional arguments
  are $x$ and $r$ but we choose the notation $\injs{\deriv{r}{x}}$ to highlight
  the definition is defined by pattern match over the various cases of the derivative operation.}
The definition is by induction on $r$.
\begin{haskell}
\injs{\deriv{\epsilon}{x}}{\_} = \{ \}
\ \ \
\injs{\deriv{\phi}{x}}{\_} = \{ \}
\ \ \
\injs{\deriv{x}{x}}{()} = \{ x \}
\ \ \
\injs{\deriv{y}{x}}{\_} = \{ \}
\\
\injs{\deriv{r^*}{x}}{(v,vs)} =  \{ v' : vs \mid v' \in \injs{\deriv{r}{x}}{v} \} 
\\
\injs{\deriv{(r_1 \conc r_2)}{x}}{} = 
 \hsbody{\lambda v.
  \hsalign{
     \CASE\ v ~\OF \\
    (v_1,v_2) \arrow \{ (v, v_2) \mid v \in \injs{\deriv{r_1}{x}}{v_1} \} \\
    \Left ~(v_1,v_2) \arrow \{ (v, v_2) \mid v \in \injs{\deriv{r_1}{x}}{v_1} \} \\
    \Right ~v_2 \arrow \{ (v,v') \mid v \in \allEps{r_1} \wedge v' \in \injs{\deriv{r_2}{x}}{v_2} \} 
  }} \\
\injs{\deriv{(r_1+r_2)}{x}}{} = 
  \hsbody{\lambda v.
  \hsalign{
     \CASE\ v ~\OF \\
     \Left \ v_1 \arrow \{ \Left\ v \mid v \in \injs{\deriv{r_1}{x}}{v_1} \} \\
     \Right  \ v_2 \arrow \{ \Right\ v \mid v \in \injs{\deriv{r_2}{x}}{v_2} \}
  }} 
\end{haskell}
\end{definition}

In the above, we use Haskell style syntax such as lambda-bound functions etc.
The first couple of cases are straightforward. For brevity,
we use the `don't care' pattern $\_$ and make use of a non-linear pattern in the third equation.
In case of Kleene star, the parse tree is represented by a sequence.
We call the injection function of the underlying expression on the first element.
In case of concatenation $r_1 \conc r_2$, we observe the shape of the parse tree
of $\deriv{r_1 \conc r_2}{x}$.  
For example, if we encounter $\Right  \ v_2$, the left component $r_1$ must be nullable.
Hence, we apply $\allEps{r_1}$.

Via a straightforward inductive proof on $r$, we can verify that the
injection function yields valid parse trees.

\begin{proposition}[Soundness of Injection]
\label{prop:injs-transform-soundness}
Let $r$ be an expression, $x$ be a symbol and $v$ be a parse tree such that
$\turns v : \deriv{r}{x}$.
Then, for any $v' \in \injs{\deriv{r}{x}}$
we find that $\turns v' : r$.
\end{proposition}

\begin{example}
  Consider our running example where $\turns \Right \ (\Right\ (), []) : \deriv{\deriv{(x+y)^*}{x}}{y}$.
  Then, 
  $\injs{\deriv{\deriv{(x+y)^*}{x}}{y}} \ (\Right \ (\Right\ (), [])) = \{ (\Left\ (), [y]) \}$
  where $\turns (\Left\ (), [y]) : \deriv{(x+y)^*}{x}$
   and $\deriv{(x+y)^*}{x} = (\epsilon + \phi) \conc (x+y)^*$.
\end{example}

As in case of $\allEps{r}$, we can only guarantee completeness for non-problematic
expressions.

\begin{proposition}[Completeness of Non-Problematic Injection]
\label{prop:injs-transform-completeness}
Let $r$ be a non-problematic expression and $v$ a parse tree such that
$\turns v : r$ where $\flatten{v} = x \conc w$ for some letter $x$ and word $w$.
Then, there exists a parse tree $v'$ such that
(1) $\turns v' : \deriv{r}{x}$ and (2) $v \in \injs{\deriv{r}{x}}{v'}$.
\end{proposition}

\begin{definition}[Parse Tree Construction]
Let $r$ be an expression.
Then, the derivative-based procedure to compute all parse trees is as follows.
\begin{haskell}
\allParse{r}{\epsilon} = \allEps{r}
\\
\allParse{r}{x \conc w} = \{ v \mid v \in \injs{\deriv{r}{x}}{v'} \wedge v' \in (\allParse{\deriv{r}{x}}{w}) \}
\end{haskell}
\end{definition}


\begin{proposition}[Valid Parse Trees]
\label{prop:valid-parse-trees}
Let $r$ be an  expression. Then, for each
$v \in \allParse{r}{\flatten{v}}$ we find that $\turns v : r$.
\end{proposition}
For non-problematic expressions, we obtain a complete parse tree construction
method.

\begin{proposition}[All Non-Problematic Parse Trees]
\label{prop:non-problem-pt}
Let $r$ be a non-problematic expression and $v$ a parse tree
such that $\turns v : r$. Then, we find that
$v \in \allParse{r}{\flatten{v}}$.
\end{proposition}

In case of a fixed expression $r$,
calls to $\allParse{r}{}$ repeatedly build
the same set of derivatives.
We can be more efficient by constructing
a finite state transducer (FST) for a fixed expression $r$
where states are descendants of $r$.
The outputs are parse tree transformation functions.
This is what we will discuss next.

\mysection{Derivative-Based Finite State Transducer}
\label{sec:all-match-fsa} 

The natural candidate for FST states are derivatives. 
That is, $\delta(r,x) = \deriv{r}{x}$.
In general, descendants (derivatives) are not finite.
Thankfully, Brzozowski showed that the set of dissimilar 
descendants is finite.

\begin{definition}[Similarity]
\label{def:similarity}
We say two expressions $r$ and $s$ are \emph{similar},
written $r \approx s$, if one can be transformed into the
other by application of the following rules.
\bda{c}
 \tlabel{Idemp} ~ r + r \approx r
\ \ \ \
 \tlabel{Comm} ~ r_1 + r_2 \approx r_2 + r_1
\\
 \tlabel{Assoc} ~ (r_1 + r_2) + r_3 \approx r_1 + (r_2 + r_3)
 \ \ \ \ 
 \tlabel{Ctxt} ~ \myirule{s \approx t}
                            {R[s] \approx R[t]}
\eda
The \tlabel{Ctxt} rules assumes expressions with a hole.
We write $R[s]$ to denote the expression where the hole  $[]$
is replaced by $s$.
\bda{c}
 \tlabel{Hole Expressions} ~
   R[] ::= [] \mid R[] \conc s \mid s \conc R[] \mid
           R[] + s \mid s + R[]
           \eda
There is no hole inside Kleene star
because the derivative operation
will only ever be applied on unfoldings of the Kleene star but never within
a Kleene star expression.

We write $\deriv{r}{}/{\approx}$ to denote the set of
equivalence classes of $\deriv{r}{}$ w.r.t.~the
equivalence relation $\approx$.
\end{definition}

\begin{proposition}[Brzozowski~\cite{321249}]
  \label{prop:Brzozowski}
$\deriv{r}{}/{\approx}$ is finite for any expression $r$.
\end{proposition}

Based on the above, we build an automata where the set of states
consists of a canonical representative for all descendants
of some expression $r$. 
A similar approach is discussed in \cite{watson93:t_axon_m}.

\begin{definition}[Canonical Representative]
  \label{def:canonical}
  For each expression $r$ we compute an expression
  $\canonic{r}$ by systematic application of the
  similarity rules:
  (1) Put alternatives in right-associative normal form
      via rule \tlabel{Assoc}.
      (2) Remove duplicates via rules \tlabel{Idemp}
      where via rule \tlabel{Comm}
      we push the right-most duplicates to the left.
      (3) Repeat until there are no further changes.
 \end{definition}

\begin{proposition}[Canonical Normal From]
  \label{prop:cnf}
  Let $r$ be an expression. Then, $\canonic{r}$
  represents a canonical normal form of $r$.
\end{proposition}

Furthermore, alternatives keep their relative position. 
For example, $\canonic{r + s + s_1 + ... + s_n + s + t} =
  r + s + s_1 + ... + s_n + t$.
  This is important for the upcoming construction
  of POSIX and Greedy parse trees.

\begin{proposition}[Finite Dissimilar Canonical Descendants]
  \label{prop:finite-dissimilar}
  Let $r$ be an expression.
  Then, the set $\Desc{r} = \{ \canonic{s} \mid r \Descendant s \}$ is finite.
\end{proposition}

Like in case of $\injs{}{}$, we need to
maintain information how to transform parse trees among similar expressions.
Hence, we attach parse tree transformation functions 
to the similarity rules.


\begin{definition}[Similarity with Parse Tree Transformation]
\label{def:similarity-transformation}
  We write $\simp{r}{f}{s}$
  to denote that expressions $r$ and $s$ are similar
  and a parse tree of $s$ can be transformed into a parse tree of $r$
  via function $f$.
  In case the function returns a set of parse trees
  we write $\simp{r}{fs}{s}$.
  We write $\simp{r}{}{s}$ if the parse tree transformation is not of interest.
\bda{c}
\tlabel{Idemp} ~
\myirule{\fs{fs} (u) = \{ \Left\ u, \Right\ u\} }
        {\simp{r + r}{fs}{r}}
\ \ \ \
\tlabel{Comm} ~
\myirule{\ba{ll}
             f (\Left\ u) & = \Right\ u
          \\ f (\Right\ u) & = \Left\ u
         \ea}
        {\simp{r_1 + r_2}{f}{r_2 + r_1}}                
\\         
\tlabel{Assoc} ~
 \myirule{\ba{ll}
           f (\Left\ u_1) & = \Left\ (\Left\ u_1)
        \\ f (\Right\ (\Left\ u_2)) & = \Left\ (\Right\ u_2)
        \\ f (\Right\ (\Right\ u_3)) & = \Right\ u_3
          \ea
         }
         {\simp{(r_1+r_2)+r_3}{f}{r_1+(r_2+r_3)}}
\ \ \ \
\tlabel{Lift} ~
\myirule{\simp{r}{f}{s}
        \\ \fs{fs} (u) = \{ f (u) \} }
        {\simp{r}{fs}{s}}
\\ 
\tlabel{C1} 
\myirule{\simp{s}{fs}{t}
          \\ \fs{gs} (u_r,u_t) = \{ (u_r, u_s) \mid u_s \in \fs{fs} (u_t) \} }
        {\simp{r \conc s}{gs}{r \conc t}}
\ \ 
\tlabel{C2} 
\myirule{\simp{s}{fs}{t}
          \\ \fs{gs} (u_t,u_r) = \{ (u_s, u_r) \mid u_s \in \fs{fs} (u_t) \} }
        {\simp{s \conc r}{gs}{t \conc r}}
\\ 
\tlabel{C3} 
\myirule{\simp{s}{fs}{t}
          \\ \ba{ll}
          \fs{gs} (\Left\ u_r) = \{ \Left\ u_r \}
       \\ \fs{gs} (\Right\ u_t) = \{ \Right\ u_s \mid u_s \in \fs{fs} (u_t) \}          
             \ea
        }
        {\simp{r+s}{gs}{r+t}}
\ \
\tlabel{C4} 
\myirule{\simp{s}{fs}{t}
          \\ \ba{ll}
          \fs{gs} (\Left\ u_t) = \{ \Left\ u_s \mid u_s \in \fs{fs} (u_t) \}     
        \\ \fs{gs} (\Right\ u_r) = \{ \Right\ u_r \}     
             \ea
        }
        {\simp{s+r}{gs}{t+r}}
\eda
\end{definition}

The above rules are derived from
the ones in Definition~\ref{def:similarity}
by providing the appropriate parse tree transformations.
Due to the similarity rule
\tlabel{Idemp} we may obtain a set of parse trees.
Rules \tlabel{C1-4} cover all the cases described
by rule \tlabel{Ctxt}.
The attached (transformation) functions yield valid parse trees (soundness)
and every parse tree of a similar expression can be obtained (completeness).

\begin{proposition}[Soundness of Transformation]
  \label{prop:sound-transformation}
Let $r$ and $s$ be two expressions and $\fs{fs}$
a function such that $\simp{r}{fs}{s}$.
Then, we find that
(1) $r \approx s$ and
(2) for any parse tree $v$ where $\turns v : s$ we have that
     $\turns v' : r$ for any $v' \in \fs{fs}(v)$.
\end{proposition}

\begin{proposition}[Completeness of Transformation]
  \label{prop:complete-transformation}
Let $r$ and $s$ be two expressions and $v$ be a parse tree
such that $\turns v : r$ and $r \approx s$.
Then, $\simp{r}{fs}{s}$ 
where $v \in \fs{fs}(v')$ for some $v'$ such that
$\turns v' : s$.
\end{proposition}

The FST to compute parse trees
for some expression $r$ consists of states $\Desc{r}$.
Each state transition from $s$ to $\canonic{\deriv{s}{x}}$
yields as output a parse tree transformer function
which is a composition of $\injs{\deriv{s}{x}}$ and $\fs{fs}$
where $\simp{\deriv{s}{x}}{fs}{\canonic{\deriv{s}{x}}}$.

\begin{definition}[FST Construction]
Let $r$ be an expression.
We define $\AllParseFST{r} = (Q,\Sigma,\delta,q_0,F)$
where $Q = \Desc{r}$, $q_0 = r$, $F= \{ s \in Q \mid \epsilon \in \lang(s) \}$
and for each $s \in Q$ and $x \in \Sigma$
we set $\delta(s,x) = (\canonic{\deriv{s}{x}},\fs{gs})$
where $\simp{\deriv{s}{x}}{fs}{\canonic{\deriv{s}{x}}}$
and $\fs{gs} (u) = \{ u_2 \mid  u_1 \in \fs{fs} (u) \wedge u_2 \in \injs{\deriv{s}{x}}{u_1} \}$.

The transition relation $\delta$ is inductively
extended to words as follows.
We define $\delta(s,\epsilon) = (s, \lambda u.\{ u \})$
and $\delta(s,x \conc w) = (r,\fs{fs})$
where $\delta(s,x) = (t,\fs{gs})$
and $\delta(t,w) = (r,\fs{hs})$
where $\fs{fs} (u) = \{ u_2  \mid u_1 \in \fs{hs} (u) \wedge u_2 \in \fs{gs} (u_1) \}$.
\end{definition}

\begin{proposition}[All Non-Problematic Parse Trees via FST]
  \label{prop:parse-fsa}
Let $r$ be a non-problematic expression and $v$ a parse tree
such that $\turns v : r$. 
Let $\AllParseFST{r} = (Q,\Sigma,\delta,q_0,F)$.
Then, we find that $v \in \fs{fs} (\allEps{r'})$
where $\delta(r,\flatten{v}) = (r',\fs{fs})$.
\end{proposition}

\mysection{Computing POSIX and Greedy Parse Trees}
\label{sec:posix-and-greedy}

Based on our earlier work~\cite{DBLP:conf/flops/SulzmannL14}
we can immediately conclude that the `first' (left-most) match obtained
by executing $\AllParseFST{r}$ is the
POSIX match.~\footnote{Technically,
  we  treat the set of parse trees like a list.
Recall that $\allEps{\cdot}$
and the simplification rule \tlabel{Idemp} favor the left-most match.
Alternatives keep their relative position in an expression.}
The use of derivatives guarantees that
the longest left-most (POSIX) parse tree is computed.

\begin{proposition}[POSIX]
  \label{prop:posix-match}
  Let $r$ be an expression and $w$ be a word
  such that $w \in \lang(r)$.
  Let $\AllParseFST{r} = (Q,\Sigma,\delta,q_0,F)$.
  Let $\delta(r,w) = (r',\fs{fs})$ for some expression $r'$
    and transformer $\fs{fs}$.
    Then, $\fs{fs} (\allEps{r'}) = \{ v_1, ..., v_n \}$
    for some parse trees $v_i$
    where $v_1$ is the POSIX match.
\end{proposition}

With little effort it is possible to customize
our FST construction to compute Greedy parse trees.
The insight is to normalize derivatives
such that they effectively correspond
to partial derivatives. Via this normalization step,
we obtain  as the `first' result
the Greedy (left-most) parse tree.
This follows from our earlier work~\cite{DBLP:conf/ppdp/SulzmannL12}
where we showed that partial derivatives naturally yield greedy matches.

We first define partial derivatives which
are a non-deterministic generalization
of derivatives. Instead of a single expression, the partial
derivative operation yields a set of expressions.

\begin{definition}[Partial Derivatives]~\footnote{We omit the smart constructor
    found in \cite{Antimirov96Partial} as this is not relevant here.}
  Let $r$ be an expression and $x$ be a symbol.
  Then, the \emph{partial derivative} of $r$ w.r.t.~$x$ is computed
  as follows:
  \bda{c}

  \ba{cc}
   \ba{l}
    \pderiv{\phi}{x}  =  \{\}
    \\ \pderiv{\epsilon}{x}  =  \{\}
   \ea
   &
   \ba{l}
   \pderiv{y}{x}  =  \left \{   
   \ba{ll}
        \{\epsilon\} & \mbox{if~$x = y$}
    \\ \{\} & \mbox{otherwise}
    \ea  \right.
    \ea
  \ea
\\  
\pderiv{r_1 + r_2}{x}  =  \pderiv{r_1}{x} \cup \pderiv{r_2}{x}
\ \ \ \
\pderiv{r^*}{x}  =  \{ r' \conc r^* \mid r' \in \pderiv{r}{x} \}
\\
   \pderiv{r_1 \conc r_2}{x} = 
     \left \{
        \ba{ll}
           \{ r_1' \conc r_2 \mid r_1' \in \pderiv{r_1}{x} \} \cup \pderiv{r_2}{x}  & 
               \mbox{if $\epsilon\in \lang(r_1)$}
         \\ \{ r_1' \conc r_2 \mid r_1' \in \pderiv{r_1}{x} \}  & \mbox{otherwise}
        \ea \right.
\eda


Let $M = \{ r_1, ..., r_n \}$ be a set of expressions.
Then, we define $+ M = r_1 + ... + r_n$
and $+ \{ \} = \phi$. 
\end{definition}

To derive partial derivatives via derivatives,
we impose the following additional similarity rules.

\begin{definition}[Partial Derivative Similarity Rules]
  \label{def:pd-sim-rules}
  \bda{c}
  \tlabel{Dist} \
  \myirule{ \ba{ll}
                 f (\Left\ (u_r, u_t)) & = (\Left\ u_r, u_t)
              \\ f (\Right\ (u_s, t_t)) & = (\Right\ u_s, u_t) 
            \ea }
          {\simp{(r + s) \conc t}{f}{r\conc t + s \conc t}}
 \ \ \
 \ba{c}
  \tlabel{ElimPhi1} \
  \simp{\phi + r}{\lambda x. \Right\ x}{r}
  \\ \\
 \tlabel{ElimPhi2} \
 \simp{\phi \conc r}{\bot}{\phi}
 \ea
  \eda
\end{definition}

Rule \tlabel{Dist} mimics the set-based operations performed
by $\pderiv{\cdot}{\cdot}$ in case of concatenation
and Kleene star. Rules \tlabel{ElimPhi1-2} cover cases
where the set is empty. We use $\bot$ to denote the undefined
parse tree transformer function.
As there is no parse tree for $\phi$ this function will
never be called.

\begin{proposition}[Partial Derivatives as Normalized Derivatives]
  \label{prop:partial-derivatives-via-derivatives}
  Let $r$ be an expression and $x$ be a symbol.
  Then, we have that $+ \pderiv{r}{x}$ is syntactically equal to
  some expression $s$ such that $\simp{\deriv{r}{x}}{}{s}$.
  We  ignore the transformer function which is not
  relevant here.
\end{proposition}

Based on the above and our earlier results
in~\cite{DBLP:conf/ppdp/SulzmannL12} we can immediately
conclude the following.

\begin{proposition}[Greedy]
  \label{prop:greedy-match}
  Let $r$ be an expression and $w$ be a word
  such that $w \in \lang(r)$.
  Let $\AllParseFST{r} = (Q,\Sigma,\delta,q_0,F)$
  where we additionally apply the similarity rules in Definition~\ref{def:pd-sim-rules}
  such that canonical representatives satisfy
  the property stated in
  Proposition \ref{prop:partial-derivatives-via-derivatives}.
  Let $\delta(r,w) = (r',\fs{fs})$ for some expression $r'$
    and transformer $\fs{fs}$.
    Then, $\fs{fs} (\allEps{r'}) = \{ v_1, ..., v_n \}$
    for some parse trees $v_i$
    where $v_1$ is the Greedy parse tree.
\end{proposition}

\begin{remark}[Linear-Time Complexity]
  Our approach has linear-time complexity~\footnote{We do not measure
    the complexity of constructing  $\AllParseFST{r}$ which can be exponential
    in the size of $r$.} in the size of the input word $w$
  assuming we treat the size of the regular expression $r$ as a constant
  and consider computation of the 'first' parse tree only.
  The size of dissimilar derivatives is at most exponential
  in the size of $r$. The size of a parse tree is bound by $r$.
  Time complexity of parse tree transformation functions is linear
  in the size of the input.
\end{remark}

There is quite a bit of scope to improve the performance
by for example employing more efficient representations
of our parse tree transformation functions.
For efficiency reasons, we also may want to specialize
$\AllParseFST{r}$ to compute the POSIX and Greedy parse trees only.
Currently, we rely on Haskell's lazy evaluation strategy
to do only the necessary work when
extracting the first parse tree from the final result
obtained by  running $\AllParseFST{r}$.
These are topics to consider in future work.


\mysection{Ambiguity Diagnosis} 
\label{sec:diagnos-ambig}

We can identify three situations where ambiguity of $r$ arises
during the construction of $\AllParseFST{r}$.
The first situation concerns nullable expressions.
If we encounter multiple empty parse trees for a nullable descendant
(accepting state) then we end up with multiple parse trees
for the initial state. Then, the initial expression is ambiguous.

The second situation concerns the case of
injecting a symbol into the parse tree of a descendant.
Recall that the $\injs{}{}$ function
from Definition~\ref{def:injection} possibly yields
a set of parse trees. This will only happen
if we apply the derivative operation
on some subterm $t_1 \conc t_2$
where $t_1$ is a nullable expression with multiple
empty parse trees. 

The final (third) situation ambiguous situation
arises in case we build canonical representatives.
Recall Definition~\ref{def:similarity-transformation}.
We end up with multiple parse trees whenever we apply
rule \tlabel{Idemp}. 

These are the only situations which may give rise to multiple parse trees.
That is, if none of these situations arises the
expression must be unambiguous.
We summarize these observations in the following result.

\begin{definition}[Realizable State]
We say that $s \in \Desc{r}$ is \emph{realizable},
if there exists a path in $\AllParseFST{r}$
such that (1) we reach $s$ and (2) along this path
all states (expressions) including $s$
do \emph{not} describe the empty language.
\end{definition}

\begin{proposition}[Ambiguity Criteria]
 \label{prop:ambiguity-criteria}
Let $r$ be a non-problematic expression.
Then, $r$ is ambiguous iff
there exists a realizable $s \in \Desc{r}$ and some
symbol $x$ where
one of the following conditions applies:
\begin{description}
\item[A1] 
          $|\allEps{s}| > 1$, or 
 \item[A2] $s=R[t_1 \conc t_2]$ where 
          $|\allEps{t_1}| > 1$, or 
 \item[A3] $\lang(\canonic{\deriv{s}{x}}) \not= \{\}$ and 
          $\simp{\deriv{s}{x}}{fs}{\canonic{\deriv{s}{x}}}$
          with rule \tlabel{Idemp} applied.
\end{description}
\end{proposition}

The above criteria are easy to verify.
In terms of the FST generated,
criteria {\bf A1} is always connected to a final state
whereas criteria {\bf A2} and {\bf A3} are always connected to transitions.
Our implementation automatically generates
the FST annotated with ambiguity information.

\begin{figure}
  \includegraphics[scale=0.7]{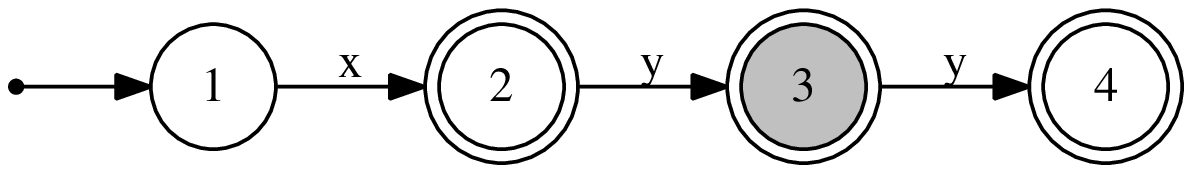}
  \caption{$\AllParseFST{(x + x \conc y) \conc (y + \epsilon)}$}
    \label{ex:exP1}
\end{figure}

In Figure~\ref{ex:exP1}, we consider the FST
for $(x + x \conc y) \conc (y + \epsilon)$.
Final state is highlighted grey to indicate that
ambiguity due to {\bf A1} arises.
Indeed, for input $x \conc y$ we can observe that
there are two distinct parse trees.
Namely, $(\Left\ x, \Left\ x)$ and $(\Right \ (x,y), \Right \ ())$.
Hence, the expression is ambiguous.

\begin{figure}
  \includegraphics[scale=0.7]{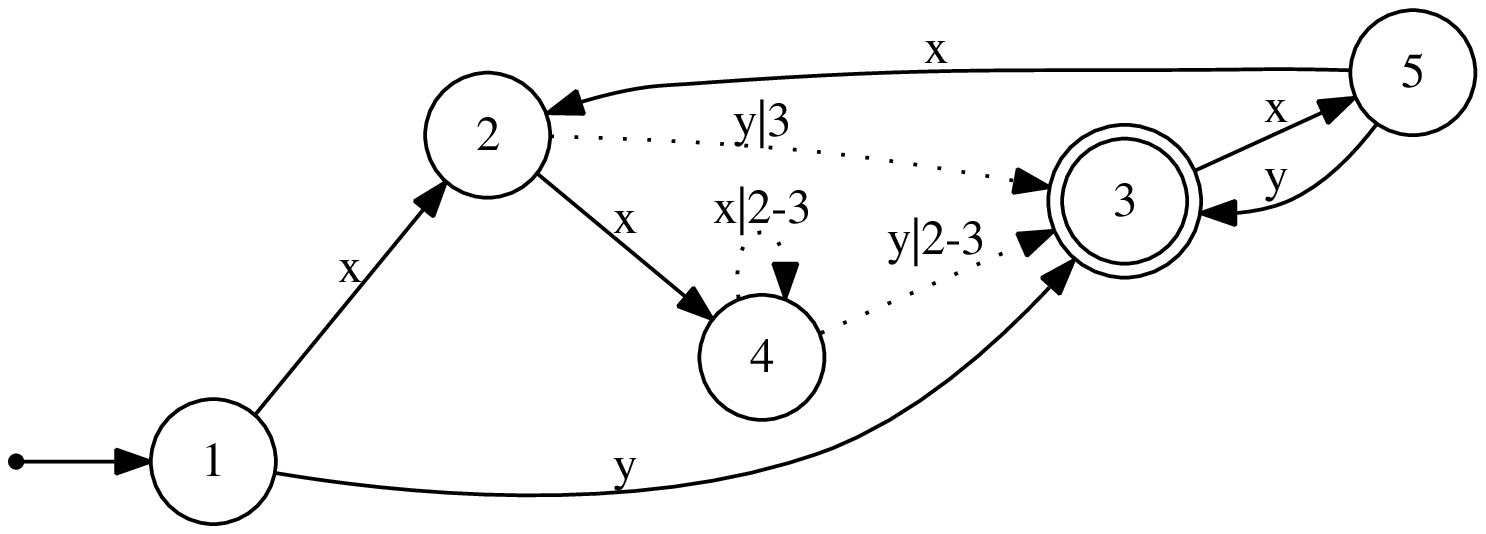}
  \caption{$\AllParseFST{(x \conc x^* + y \conc x + x \conc y \conc x)^* \conc y}$}
    \label{ex:exP3}
  \end{figure}

Consider another example taken from \cite{DBLP:conf/wia/BorsottiBCM15}.
See Figure~\ref{ex:exP3}.
We find ambiguous transitions due to {\bf A2} and {\bf A3}.
Such transitions are represented as dotted arrows
with labels to indicate {\bf A2} and {\bf A3}.
Ambiguity due to {\bf A1} does not arise for this example.

Let us investigate the ambiguous transition from state 2 to state 3.
We carry out the constructions of states starting
with the initial expression $r \conc y$
where $r=(x \conc x^* + y \conc x + x \conc y \conc x)^*$.
For brevity, we make use of additional similarity rules
such as $\epsilon \conc s \approx s$ to keep the size of descendants
manageable. In the following, we write $r \stackrel{x}{\rightarrow} s$
if $s=\deriv{r}{x}$.

\bda{ll}
     & r \conc y
\\ \stackrel{x}{\rightarrow} & ((x^* + y \conc x) \conc r) \conc y
\\ \stackrel{y}{\rightarrow} & (x \conc r) \conc y + (x \conc r) \conc y + \epsilon
\\ \approx  &  (x \conc r) \conc y + \epsilon
\eda

In the last step, we apply rule \tlabel{Idemp}.
Hence, the ambiguous transition from state 2 to state 3.

State 3 is final, however, $x \conc y$ is not yet a full counter-example
to exhibit ambiguity. In essence, $x \conc y$ is a prefix of the
full counter-example $x \conc y \conc x \conc y$.
For this example, we obtain parse trees
$([\Left \ (x,[]), \Left \ (\Right \ (x,y))], y)$
and $(\Right \ (\Right \ (x,(y,x))), y)$.
The first one is obtained via Greedy
and the second one via POSIX.

To summarize, from the FST it is straightforward to
derive minimal prefixes of counter-examples.
To obtain actual counter-examples, minimal prefixes
need to be extended so that a final state is reached.
Based on the FST, we could perform a breadth-first search
to calculate all such minimal counter-examples.
Alternatively, we can built (minimal) counter-examples during the
construction of the FST.

There is clearly much scope for more sophisticated ambiguity diagnosis
based on the information provided by the FST.
An immediate application is to check (statically)
any differences among Greedy and POSIX.
We simply check both methods against the set of
minimal counter-examples.
It is clear that there are only finitely many (minimal) counter-examples
as there are a finite number of states and transitions.
Obtaining more precise bounds on their size is something to consider
in future work.

\bibliography{main}

\newpage

\appendix

\section{Proof Sketches}

\subsection{Propositions \ref{prop:empty-parse-trees} and \ref{prop:allempty}}

For both statements, we proceed by induction on $r$.
Consider Proposition~\ref{prop:allempty}.
For case $r^*$ due to the 'non-problematic' assumption
we find that $\turns [] : r^*$ is the only possible candidate.

\subsection{Propositions \ref{prop:injs-transform-soundness} and \ref{prop:injs-transform-completeness}}

Again by induction on $r$.
For Proposition \ref{prop:injs-transform-completeness},
we consider some of the  interesting cases.
 
 \begin{itemize}
  \item Case $r^*$ where $\turns v : r^*$:
     \begin{enumerate}
       \item By assumption $r^*$ is non-problematic.
             Hence, $v = v_1:vs$ for some $v_1$ and $vs$
             where $\turns v_1 : r$ and $\flatten{v_1} = x \conc w$ for some word $w$.
      \item By induction there exists $v'$ such that
            $\turns v' : \deriv{r}{x}$ and $v_1 \in \injs{\deriv{r}{x}}{v'}$.
      \item By definition of $\injs{}{}$ we find that
            $v_1:vs \in \injs{\deriv{r^*}{x}}{v':vs}$.
      \item Hence, there exists $v':vs$ which guarantees (1) and (2)
            and we are done.           
     \end{enumerate}
  \item Case $r_1 \conc r_2$ where $\turns v : r_1 \conc r_2$:
     \begin{enumerate}
       \item By assumption $v= (v_1,v_2)$ where
             $\turns v_1 : r_1$, $\turns v_2 : r_2$ and $\flatten{v_1} = x \conc w$
             for some word $w$.
      \item Suppose $\epsilon \not \in \lang(r_1)$.
         \begin{enumerate}
           \item By induction there exists $v'$ such that
                 $\turns v' : \deriv{r_1}{x}$ and $v_1 \in \injs{\deriv{r_1}{x}}{v'}$.
           \item Under our assumptions $\deriv{r_1 \conc r_2}{x} = \deriv{r_1}{x} \conc r_2$.
           \item By definition of $\injs{}{}$ we find that
                 $(v_1,v_2) \in \injs{\deriv{r_1 \conc r_2}{x}}{(v',v_2)}$.
           \item Hence, there exists $(v',v_2)$ which guarantees (1) and (2)
            and we are done.           
         \end{enumerate}
      \item Otherwise $\epsilon \in \lang(r_1)$ where we assume that
              $v_1 = x \conc w$ for some word $w$:
       \begin{enumerate}
         \item Similar to the above reasoning we find
               $(v_1,v_2) \in \injs{\deriv{r_1}{x} \conc r_2}{(v',v_2)}$
               for some $v'$.
         \item By assumption $\deriv{r_1 \conc r_2}{x} = \deriv{r_1}{x} \conc r_2 + \deriv{r_2}{x}$.
         \item Hence,  $(v_1,v_2) \in \injs{\deriv{r_1 \conc r_2}{x}}{\Left\ (v',v_2)}$
               via which we can establish (1) and (2) and we are done.
       \end{enumerate}
      \item The only remaining case is that $\epsilon \in \lang(r_1)$
             and $\flatten{v_1} = \epsilon$.
         \begin{enumerate}
            \item Hence, $\flatten{v_2} = x \conc w$ for some word $w$.
            \item By induction on $r_2$ we find that there exists $v'$ such that
                 $\turns v' : \deriv{r_2}{x}$ and $v_2 \in \injs{\deriv{2_1}{x}}{v'}$.
            \item By Proposition~\ref{prop:allempty} we obtain that $v_1 \in \allEps{r_1}$.
            \item By definition of $\injs{}{}$ we find that
                 $(v_1,v_2) \in \injs{\deriv{r_1 \conc r_2}{x}}{\Right\ v_2}$
               which concludes the proof for the subcase of concatenated expressions.  
         \end{enumerate}
     \end{enumerate}
 \end{itemize}

\subsection{Propositions  \ref{prop:valid-parse-trees} and \ref{prop:non-problem-pt}}

Proposition  \ref{prop:valid-parse-trees}
follows straightforwardly
from Propositions \ref{prop:empty-parse-trees} and \ref{prop:injs-transform-soundness}.

Consider Proposition \ref{prop:non-problem-pt} where we need
to verify that for all non-problematic $r$ and $v$ where $\turns v : r$
we find that $v \in \allParse{r}{\flatten{v}}$.
We proceed by  induction on $\flatten{v}$.

{\bf Case} $\flatten{v} = \epsilon$:
By Proposition \ref{prop:allempty} we find that $v \in \allEps{r}$.
By definition of $\allParse{}{}$ we conclude that $v \in \allParse{r}{\flatten{v}}$ and we are done.

{\bf Case} $\flatten{v} = x \conc w$:
By Proposition \ref{prop:injs-transform-completeness},
there exists $v'$ such that $\turns v' : \deriv{r}{x}$, $v \in \injs{\deriv{r}{x}}{v'}$
and $\flatten{v'} = w$.
Via a simple induction we can verify that if $r$ is non-problematic so must be $\deriv{r}{x}$.
By I.H. (for $\deriv{r}{x}$ and $v'$) we find that $v' \in \allParse{\deriv{r}{x}}{w}$.
By definition of $\allParse{}{}$ we conclude that $v \in \allParse{r}{\flatten{v}}$ and we are done.

\subsection{Propositions \ref{prop:cnf} and \ref{prop:finite-dissimilar}}

Consider Proposition~\ref{prop:cnf}.
We can show that the thus systematically applied similarity rules
represent a terminating and confluent rewrite system.
Hence, we obtain canonical normal forms.

Consider Proposition\ref{prop:finite-dissimilar}.
The set of canonical representatives is finite.
Follows from Brzozowski's result Proposition \ref{prop:Brzozowski}.

\subsection{Propositions \ref{prop:sound-transformation} and \ref{prop:complete-transformation}}

Both results follow by induction.
For Proposition \ref{prop:sound-transformation},
by induction on the derivation $\simp{r}{fs}{s}$.
For Proposition \ref{prop:complete-transformation},
by induction $r \approx s$.

\subsection{Proposition \ref{prop:parse-fsa}}

The set of states and transitions is finite.
Follows from Proposition \ref{prop:finite-dissimilar}.
The resulting parse trees must valid as this follows
from the respective results for $\injs{}{}$
and parse tree transformations resulting from similarity.
The same applies for the completeness direction where
we require the assumption that the expression
is non-problematic.

\subsection{Proposition \ref{prop:posix-match}}

Follows from our earlier results stated in \cite{DBLP:conf/flops/SulzmannL14}.
Note that we strictly favor left-most parse trees.
Recall the definitions
\bda{c}
\allEps{ r_1+r_2 } =  \{ \Left\ v_1 \mid v_1 \in \allEps{r_1} \}
                      \cup \{ \Right\ v_2 \mid v_2 \in \allEps{r_2} \}             
\eda
and
\bda{c}
\tlabel{Idemp} ~
\myirule{fs (u) = \{ \Left\ u, \Right\ u\} }
        {\simp{r + r}{fs}{r}}
\eda

\subsection{Proposition \ref{prop:partial-derivatives-via-derivatives}}

{\bf Helper statement:}
  Let $r$ and $x$ be such that $\pderiv{r}{x} = \{ \}$.
  Then, we find that $\simp{\deriv{r}{x}}{}{\sum_{t \in F} t}$
  where $F$ is a finite set consisting of terms of the shape $\phi$ or $\phi\conc t'$ for some
  expression $t'$.

  The proof of this statement is by induction on $r$.
  
  {\bf Case}s for $\phi$, $y$ and $\epsilon$ are straightforward.
  
  {\bf Case} $s^*$:
  By assumption $\pderiv{s^*}{x} = \{ \}$. Hence, $\pderiv{s}{x} = \{ \}$.
  By induction $\simp{\deriv{s}{x}}{}{\sum_{t \in F} t}$ for some $F$ as described above.
  By definition $\deriv{s^*}{x} = \deriv{s}{x} \conc s^*$.
  By similarity, we find that $\simp{\deriv{s}{x} \conc s^*}{}{(\sum_{t \in F} t) \conc s^*}$.
  By similarity rule \tlabel{Dist}, we find that
  $\simp{(\sum_{t \in F} t) \conc s^*}{}{\sum_{t \in F'} t}$
  where $F' = \{ t \conc s^* \mid t \in F \}$.
  Hence, $\simp{\deriv{s^*}{x}}{}{\sum_{t \in F'} t}$ and we are done.

  {\bf Case} $s_1 + s_2$: Straightforward by induction.

  {\bf Case} $s_1 \conc s_2$: By induction and application of similarity rule \tlabel{Dist}.

  \mbox{}
  \\
{\bf Proof of Proposition:}  
We proceed to verify Proposition \ref{prop:partial-derivatives-via-derivatives}
by induction on $r$. We assume that $=$ denotes for syntactic equality.

{\bf Case } $r_1 + r_2$:
By induction $+\pderiv{r_i}{x} = s_i$ for some $s_i$ where $\simp{\deriv{r_i}{x}}{}{s_i}$
for $i=1,2$.

{\bf Subcase1}: Suppose the sets $\pderiv{r_i}{x}$ are non-empty:
\bda{c}
 \pderiv{r_1}{x} = \{ t_1,...,t_l, t'_1,...,t'_m\}
 \ \ \ \
 \pderiv{r_2}{x} = \{ t''_1,...,t''_n, t'_1,...,t'_m\}
 \eda
 for some $t_j$, $t'_j$, $t''_j$
 where $t'_j$ describe some common parts.
 
By definition $\pderiv{r_1 + r_2}{x} = \{ t_1,...,t_l, t'_1,...,t'_m, t''_1,...,t''_n \}$.
By definition $\deriv{r_1 + r_2}{x} = \deriv{r_1}{x} + \deriv{r_2}{x}$.
Via similarity rules \tlabel{Idemp} and \tlabel{Comm}
we can remove duplicates $t'_j$.
Thus, 
we find that $\simp{\deriv{r_1 + r_2}{x}}{}{t_1 + ... + t_l + t'_1 + ... + t'_m + t''_1 + ... + t''_n}$
and  we are done.

{\bf Subcase2}: Suppose both sets $\pderiv{r_i}{x}$ are empty. Hence, $\pderiv{r_1 + r_2}{x} = \{ \}$
and $+\pderiv{r_1 + r_2}{x} = \phi$.
Via the helper statement we can conclude that $\simp{\deriv{r_1}{x}}{}{\sum_{t \in F_1} t}$
and $\simp{\deriv{r_2}{x}}{}{\sum_{t \in F_2} t}$.
Via similarity rules \tlabel{ElimPhi1-2}, we can guarantee
that $\simp{\deriv{r_1 + r_2}{x}}{x}{\phi}$ and we are done.

{\bf Subcase3}: One of the sets $\pderiv{r_i}{x}$ is empty and the other is non-empty.
Similar reasoning as above.

{\bf Case} $r_1 \conc r_2$:  We only consider the subcase where $\epsilon \not \in \lang(r_1)$
and $\pderiv{r_1}{x}$ is non-empty. Suppose $\pderiv{r_1}{x} = \{s_{1_1} , .... , s_{1_m}\}$
and therefore $+\pderiv{r_1}{x} = s_{1_1} + .... + s_{1_m}$.
By induction $\simp{\deriv{r_1}{x}}{}{s_{1_1} + .... + s_{1_m}}$.

We conclude that $+\pderiv{r_1 \conc r_2}{x} = + \{ s_{1_1} \conc r_2, ..., s_{1_m} \conc r_2 \} = 
  s_{1_1} \conc r_2 + ... + s_{1_m} \conc r_2$.

By definition, $\deriv{r_1 \conc r_2}{x} = \deriv{r_1}{x} \conc r_2$.
By similarity and the above
we find that $\simp{\deriv{r_1 \conc r_2}{x}}{}{s_{1_1} \conc r_2 + ... + s_{1_m} \conc r_2}$.

Thus, for some $s$ we have that
$+\pderiv{r_1 \conc r_2}{x} = s$ and $\simp{\deriv{r_1 \conc r_2}{x}}{}{s}$.
Take $s$ equal to $s_{1_1} \conc r_2 + ... + s_{1_m} \conc r_2$.

The other cases can be proven similarly.

\subsection{Proposition \ref{prop:greedy-match}}

Proposition \ref{prop:partial-derivatives-via-derivatives} is crucial here.
Via the additional similarity rules from Definition~\ref{def:pd-sim-rules}
we can normalize derivatives such that they effectively correspond to partial
derivatives.
The result follows then straightforwardly
from our earlier results stated in \cite{DBLP:conf/ppdp/SulzmannL12}.

\subsection{Proposition \ref{prop:ambiguity-criteria}}

If any of the criteria {\bf A1-3} arise, we can construct
a counter-example. See discussion in Section~\ref{sec:diagnos-ambig}.
Hence, the expression must be ambiguous.

If none of the criteria arises, the expression must be unambiguous.
There is no ambiguity in the parse tree construction.
Our completeness results guarantee that all possible parse trees are covered.

\end{document}